# Angular and temperature dependence of current induced spin-orbit effective fields in Ta/CoFeB/MgO nanowires

Xuepeng Qiu[1], Praveen Deorani[1], Kulothungasagaran Narayanapillai[1], Ki-Seung Lee[2,3], Kyung-Jin Lee[2,3,4], Hyun-Woo Lee[5] & Hyunsoo Yang[1]

[1]*Department of Electrical and Computer Engineering, National University of Singapore, 117576, Singapore*

[2]*Department of Materials Science and Engineering, Korea University, Seoul 136-701, Korea*

[3]*Spin Convergence Research Center, Korea Institute of Science and Technology, Seoul 136-791, Korea*

[4]*KU-KIST Graduate School of Converging Science and Technology, Korea University, Seoul 136-713, Korea*

[5]*PCTP and Department of Physics, Pohang University of Science and Technology, Kyungbuk 790-784, Korea*

Correspondence and requests for materials should be addressed to H.Y. (eleyang@nus.edu.sg)

**Current induced spin-orbit effective magnetic fields in metal/ferromagnet/oxide trilayers provide a new way to manipulate the magnetization, which is an alternative to the conventional current induced spin transfer torque arising from noncollinear magnetization. Ta/CoFeB/MgO structures are expected to be useful for non-volatile memories and logic devices due to its perpendicular anisotropy and large current induced spin-orbit effective**

**fields. However many aspects such as the angular and temperature dependent phenomena of the effective fields are little understood. Here, we evaluate the angular and temperature dependence of the current-induced spin-orbit effective fields considering contributions from both the anomalous and planar Hall effects. The longitudinal and transverse components of effective fields are found to have strong angular dependence on the magnetization direction at 300 K. The transverse field decreases significantly with decreasing temperature, whereas the longitudinal field shows weaker temperature dependence. Our results reveal important features and provide an opportunity for a more comprehensive understanding of current induced spin-orbit effective fields.**

In-plane current induced magnetization switching[1-6], fast domain wall motion[7-12] and high frequency magnetization oscillation[13-15] have been demonstrated by utilizing the newly discovered current induced spin-orbit effective fields in magnetic multilayers, especially in heavy metal/perpendicularly magnetized ferromagnet/oxide trilayers. In contrast to the conventional spin transfer torque which exploits the noncollinear magnetization in metallic spin valves and magnetic tunnel junctions[16,17], in-plane current induced spin-orbit effective fields in trilayers exploit the spin-orbit coupling to generate a torque even on the collinear homogeneous ferromagnetic layer. Since trilayers have broken structural inversion symmetry and contain a heavy metal layer, such as Pt or Ta, with a strong atomic spin-orbit coupling, both the spin Hall effect[2,3,8,10] and the Rashba spin-orbit interaction[1,7,18-21] can be relevant in terms of symmetry and are proposed as the underlying mechanisms of the effective fields. The current induced spin-orbit effective fields can be decomposed into two components[3,18,21-24] such as the longitudinal field ($H_L$) along $\hat{y} \times \hat{m}$ direction and the transverse field ($H_T$) along $\hat{y}$ direction, where $\hat{m}$ is the magnetization and $\hat{y}$ is the directional unit vector (transverse to the current and in the film plane). Recent studies show that the size and even the sign of both $H_L$ and $H_T$ depend on the heavy metal layer thickness[22] and the capping layer thickness[25], suggesting complex underlying mechanisms contributing to these effective fields. Complex angular dependences of current induced spin-orbit effective fields reported in $AlO_x$/Co/Pt and MgO/CoFeB/Ta systems emphasize the prominent role of interfaces[26]. These results suggest that current induced spin-orbit effective fields are still at the early stage of understanding. In order to achieve a better understanding of current induced effective fields, one should go beyond the symmetry perspective, and more quantitative comparison is required between experiments and theories. In this respect, the detailed angle dependent characterizations of the fields at different temperatures are of great importance.

The discovery of perpendicular anisotropy in Ta/CoFeB/MgO structures[27] is not only a novel building block for spin transfer torque based applications, but also of significance for the study and application of current induced spin-orbit effective fields. In-plane current injection in Ta/CoFe(B)/MgO structures drives higher efficient magnetization switching and domain wall motion with opposite polarity compared to that in Pt/CoFe(B)/MgO,[2,8] which indicates larger and opposite current induced spin-orbit effective fields in Ta based trilayers compared to those based in Pt. These observations are consistent with the opposite signs of spin Hall angles in Ta and Pt. However, the magnitude of the spin Hall angle in Ta is still under debate. While some group reports larger spin Hall angle for Ta than Pt,[2] some other group find otherwise[28,29]. It should be noted that both the longitudinal and transverse effective fields can exert a torque on the magnetization and influence the spin dynamic process, and each of the field components has been suggested to originate from different underlying mechanisms[22,26]. The coexistence of both the longitudinal and transverse effective fields in Ta/CoFeB/MgO offers an ideal system to study both effective field components[22,26], in contrast to Pt based trilayers, in which the transverse field is still controversial among reports with some having reported it to be very small or negligible[3,9].

Here we measure the angular and temperature dependence of current induced spin-orbit effective fields from Ta/CoFeB/MgO nanowires using a lock-in technique[22,30]. Analytical formulae have been developed in the Cartesian coordinate system to evaluate the current-induced effective fields with consideration of both the anomalous and planar Hall effects. Using vector measurements the decomposed transverse field ($H_T$) is found to be larger than the longitudinal field ($H_L$) at 300 K, and both $H_T$ and $H_L$ have a strong dependence on the magnetization direction. The respective effective field ($H_T$ or $H_L$) direction in Ta/CoFeB/MgO is opposite compared to that of Pt/CoFeB/MgO (Supplementary Information S5). Upon decreasing the temperature, the strength of both field components decreases, but their directions remain unchanged, and $H_T$ shows stronger temperature dependence than $H_L$.

## Results

The stacked film is of the form, Si substrate/Ta (2)/Co$_{40}$Fe$_{40}$B$_{20}$ (0.8)/MgO (2)/SiO$_2$ (3) (nominal thicknesses in nm) grown by high vacuum (< $2\times10^{-9}$ Torr) magnetron sputtering. The film is annealed at 170 °C for one hour to obtain good perpendicular anisotropy. The films are patterned into Hall bars with dimension 600 nm × 600 nm by e-beam lithography as shown in Fig. 1(a). The geometry of the transport measurements is shown in Fig. 1(b). Due to the perpendicular magnetic anisotropy of the Ta/CoFeB/MgO system[27], the easy axis of the magnetization in the CoFeB layer is oriented along the out-of-plane direction, as proved by the anomalous Hall loop in Fig. 1(c), where the magnetic field is applied along the out-of-plane direction. Figure 1(d) shows the current induced switching in a Ta/CoFeB/MgO nanowire with a 400 Oe external magnetic field applied along the current direction. The anomalous Hall signal ($R_H$) can be changed between a high and a low resistance states by injecting an in-plane current, indicating the magnetization switching of the CoFeB layer between the $M_z > 0$ and $M_z < 0$ states. The switching current is ~ 0.11 mA which corresponds to a current density of $6.55 \times 10^6$ A/cm$^2$ assuming the current is flowing uniformly throughout the Ta/CoFeB layer. The magnitude of switching current density is comparable to the spin transfer torque induced switching in MgO based current perpendicular to plane (CPP) magnetic tunnel junction (MTJ) nanopillars[27,31], demonstrating an alternative for future high density memories and logic devices. The abrupt magnetization switching characteristics in Fig. 1(c) and (d), either by field or current, indicate well-defined perpendicular anisotropy of the device.

**Two measurement schemes and current induced spin-orbit effective fields.** In order to quantify the current induced spin-orbit effective fields, we measure the anomalous Hall voltage loops using a harmonic technique. A sinusoidal current $I_{ac}$ with the amplitude of 28.3 μA and the frequency of 13.7 Hz is applied to the nanowires, and the first and second harmonic signals ($V_f$ and $V_{2f}$, respectively) are measured simultaneously by two lock-in amplifiers. Each data point averages over 40 readings to reduce the noise. The Joule heating effect induced by currents is negligible (Supplementary Information S4). The measurements are performed in two measurement schemes separately, such as the longitudinal and the transverse scheme as illustrated in Fig. 2(a) and 2(b), respectively. Different from the previous studies[22,30], where the magnetic field is swept below the coercive field, we sweep the magnetic field in a wider range so that the magnetization becomes saturated in the film plane for the large sweeping fields. In both schemes, the external magnetic field is applied with a small tilting angle $\theta_H$ to the film plane. In the longitudinal scheme in Fig. 2(a), both the ac current and magnetic field are applied along the *x*-direction. The anomalous Hall loops measured in the longitudinal scheme are shown in Fig. 2(c). With field sweeps along the current flow direction, the second harmonic signal $V_{2f,//}$ shows a negative peak at a positive magnetic field and a positive peak at a negative magnetic field. For the transverse scheme, the magnetic field is applied along the *y*-direction perpendicular to the current flow, as shown in Fig. 2(b). Figure 2(d) shows the anomalous Hall loops in the transverse configuration, in which the second harmonic signal $V_{2f,\perp}$ shows a negative peak for both polarities of the magnetic field.

The second harmonic signal $V_{2f}$ is induced by the ac current of frequency *f* which exerts a periodic effective field on the magnetization, so that the magnetization oscillates around the equilibrium direction with an oscillation amplitude of $\Delta\theta$ and $\Delta\phi$. Combined with the anomalous Hall effect (AHE), which depends on $\theta$, and with the planar Hall effect (PHE), which depends both on $\theta$ and $\phi$, the oscillation generates the second harmonic Hall voltage $V_{2f}$. In the Cartesian coordinate system as defined in Fig. 1(b), the relationship between the second harmonic voltage ($V_{2f,//}$ and $V_{2f,\perp}$), magnetization direction ($\theta$ and $\phi$), and oscillation amplitude ($\Delta\theta$ and $\Delta\phi$) can be written as (Supplementary Information S1)

$$\begin{aligned}V_{2f,//} &= -\frac{I_{ac}R_{AHE}\cos\theta\Delta\theta}{2} - I_{ac}R_{PHE}\cos^2\theta\Delta\phi, \text{ (H in } x\text{ - }z \text{ plane, } \phi = 0°) \\ V_{2f,\perp} &= -\frac{I_{ac}R_{AHE}\cos\theta\Delta\theta}{2} + I_{ac}R_{PHE}\cos^2\theta\Delta\phi, \text{ (H in } y\text{ - }z \text{ plane, } \phi = 90°).\end{aligned} \quad (1)$$

Since the magnetization oscillations ($\Delta\theta$, $\Delta\phi$) are induced by current induced effective fields, the values of $\Delta\theta$ and $\Delta\phi$ are determined by the strengths of effective fields, $H_L$ and $H_T$. The relationships among $\theta$, $\phi$, $\Delta\theta$, $\Delta\phi$, $H_L$, and $H_T$ are obtained by solving the force balance equations (Supplementary Information S2).

With the formulae derived above, we first attempt to fit the $V_{2f}$ curves by assuming constant values of $H_L$ and $H_T$ independent of the magnetization direction. This assumption will be removed later to examine the angular dependence of the effective fields. Figures 2(e, f) show the best fits under this assumption with known device parameters of $R_{AHE}$ = 7.42 Ω and $R_{PHE}$ = 0.43 Ω. Values of $H_L$ = -7.4 Oe, $H_T$ = -32.7 Oe, and $H_{an}$ = 4550 Oe are obtained, which correspond to $H_L$ = -440 Oe and $H_T$ = -1942 Oe at $10^8$ A/cm$^2$ (see Supplementary S3 for the detailed fitting procedures). The directions of $H_L$ and $H_T$ are consistent with previous reports[2,6,8,26]. The influence of planar Hall effect is important as discussed previously[26] and also in Supplementary Information S7, since its neglect results in underestimation of $H_L$ and $H_T$. The maximum Oersted field generated from the applied ac current is about 0.2 Oe in the opposite direction to $H_T$. The anomalous Nernst effect (ANE) is also evaluated[26] (see Supplementary Information S8). The contributions from the Oersted field and ANE are negligibly small and excluded from our evaluation of the fields. It should be noted that the ratio of $|H_L/H_T|=0.23$ is similar to the previous reports from a Ta/CoFeB/MgO structure[22] and Co/Pd multilayers[23], but quite different from a previous study[26]. Away from the dip, the fitted curve, especially the one for $V_{2f,\perp}$ in Fig. 2(f), deviates from the experimental data. The deviation suggests that, contrary to the assumption, the current induced spin-orbit effective fields have an angular dependence on the magnetization direction.

**Angular and temperature dependence of current induced spin-orbit effective fields.** The previous studies have tried the linear fitting method with assuming angle-independent $H_L$ and $H_T$ for the high $\theta$ range[22,30]. In order to study the angular dependence of current induced spin-orbit effective fields, we calculate the effective fields at each magnetic field (see Supplementary S3 for the detailed calculation procedures). With the conversion of the magnetic field into $\theta$ using the first harmonic Hall voltage, we obtain the angular dependence of current induced effective fields at 300 K, as shown in Fig. 3(a, b). Indeed both $H_L$ and $H_T$ depend on $\theta$ and are consistent with a recent report[26], which also showed strong angular dependence of both effective fields. It should be noted that the studied range of $\theta$ (10 ~ 80°) in our work is wider than the previous study (30 ~ 80° in our geometry). An interesting observation is that when the magnetization lies toward the plane of a ferromagnet layer ($\theta$ = 10 ~ 30°), both $H_L$ and $H_T$ increase more rapidly and show different angular dependence as compared to a monotonic dependence in the high $\theta$ range (40 ~ 80°). This suggests the effective fields have different angular dependence at low and high angle regions.

The harmonic loops are also measured at different temperatures in order to study the temperature dependence of $H_L$ and $H_T$. First we adopt the same strategy as in Fig. 2(e, f) by assuming constant values of $H_L$ and $H_T$ to evaluate the effective fields at different temperatures. Variations of $R_{AHE}$, $R_{PHE}$, and $H_{an}$ with temperature are considered (see Supplementary Information S3). The results are shown in Fig. 4(a). $H_L$ changes from -7.4 Oe at 300 K to -4.9 Oe at 50 K, while $H_T$ shows a more rapid decrease from -32.7 Oe at 300 K to -2.77 Oe at 50 K. We compare the temperature dependence of $H_L$ with the current induced switching current $I_S$, since current induced switching is mainly driven by $H_L$.[1,3,32] As seen from Fig. 4(b), the temperature dependence of $H_L/H_{an}$ shows a similar trend with $1/I_S$. When temperature changes from 300 to 50

K, $H_L/H_{an}$ decreases 3.6 times and $1/I_S$ decrease 4.8 times, whereas $H_T/H_{an}$ in the inset of Fig. 4(b) shows a change more than one order within the temperature range. The small difference between the change of $H_L/H_{an}$ and $1/I_S$ with temperature indicates $H_T$ may also play a role in the current induced switching.

Angular dependence of $H_L$ and $H_T$ at different temperatures are also shown in Fig. 4(c, d). Fluctuations at large $\theta$ in Fig. 4(c), especially at low temperatures (50 and 100 K), can be attributed to the small second harmonic signals, when magnetization is close to the out-of-plane direction at low fields. Overall both $H_L$ and $H_T$ are found to decrease with decreasing temperature at different $\theta$, and a stronger dependence on $\theta$ is observed at higher temperatures. An interesting observation is that $H_L$ becomes angular independent for the studied range of $\theta$ at 50 and 100 K, but shows a weak angular dependence for $\theta < 15°$ at 150 K. In the inset of Fig. 4(c), a fit for the $V_{2f,//}$ data (mainly originates from $H_L$) at 100 K by assuming field-independent effective fields shows good match at each magnetic field, verifying $H_L$ is angular independent at 100 K. On the other hand, the fitting to the experimental $V_{2f,\perp}$ data at 100 K shows large deviation as shown in the inset of Fig. 4(d), suggesting $H_T$ is angular dependent at 100 K although its magnitude becomes much smaller compared to that at 300 K. A theoretical model is required to explain the observed complex temperature and angular dependence in current induced spin-orbit effective fields.

**Discussion**

Quite different temperature dependence of $H_L$ and $H_T$ may indicate the possibility of different origins for these two effective field components. Indeed, $H_L$ is often argued to arise mainly from the spin Hall effect[3,8,33,34], while $H_T$ mainly from the Rashba interaction effect[1,6,7,18,21]. Assuming the spin Hall effect is the origin of $H_L$, the relation between $H_L$ and the spin Hall angle $\theta_{SH}$ is given by $H_L = \hbar \theta_{SH} |j_e|/(2|e|M_S t_F)$, where $e$ is the electron charge, $M_S = 10^6$ A/m is the saturation magnetization, and $t_F = 0.8$ nm is the ferromagnet thickness[24,35]. By using the $H_L$ value of -7.4 Oe (corresponding to -440 Oe per $10^8$ A/cm$^2$) in Fig. 2(e, f), $\theta_{SH}$ of Ta is estimated to be -0.11 at 300 K which is close to the reported value of -0.12.[2] However, as the Ta thickness ($t_{Ta}$) is 2 nm in our structure, the spin diffusion term $(1-\mathrm{sech}(t_{Ta}/\lambda_{Ta}))$ should be considered to extract $\theta_{SH}$, which enlarges $\theta_{SH}$ of Ta to -0.26 by using $\lambda_{Ta} = 1.8$ nm from a recent report[36]. The small decrease of $H_L$ as temperature decreases can be related to the enhancement of $M_S$ and small variation of $\theta_{SH}$ at lower temperatures[37]. However, the currently established spin Hall model cannot explain the angular dependence of $H_L$. Therefore, either an extended spin Hall scenario or another mechanism is required to fully describe the temperature and angular dependence of $H_L$. Similarly, the features of $H_T$ in this study cannot be explained by the currently established Rashba interaction model, which assumes the transverse field to be angle-independent[18,33] or weakly dependent[38]. Thus it fails to describe the observed angular dependence of $H_T$. The temperature dependence of $H_T$ also presents challenge for theories, since $H_T$ decreases significantly with lowering temperature and becomes negligible at 50 K, suggesting the possibility that it may be related thermally induced excitation process[39]. If the Rashba interaction is considered to be responsible for $H_T$, our observations call for an extension or modification of the model to include a temperature dependence, which lacks at this moment. Recently it has been proposed that the imaginary part of the spin mixing conductance can result in $H_T$.[40] However, as the spin mixing conductance is little influenced by temperature[41], the spin mixing conductance scenario fails to explain the temperature dependence of $H_T$. Nonetheless, since $H_T$ is a few times larger than $H_L$ at room temperature and increases more rapidly with increasing temperature, $H_T$ can play an important role for practical device applications. Recently, $H_T$ has been reported from in-plane magnetization Pt/NiFe structures and utilized for magnetization switching[4].

In conclusion, we report strong angle and temperature dependence of in-plane current induced spin-orbit effective fields. An analytical framework is developed in the Cartesian coordinate system to evaluate current induced spin-orbit effective fields from low frequency harmonic Hall voltage measurements. Our results illustrate the complex characteristics of the spin-orbit effective magnetic fields depending on the direction of magnetization and temperature. However, these results cannot be completely explained by either the spin Hall or Rashba interaction models, and call for a new model of current induced effective fields. As the origin of the effective fields is still under intense debate, the angle and temperature dependent characterization of these fields reveals detailed features of the current induced effective fields in metal/ferromagnet/oxide trilayers, and allows optimizing the effect for future spin-orbit based spintronic devices.

**Acknowledgements**

This work is partially supported by the Singapore National Research Foundation under CRP Award No. NRF-CRP 4-2008-06, NRF (NRF-2013R1A2A2A01013188), KU-KIST School Joint Research Program, NRF (NRF-2013R1A2A2A05006237), and MOTIE (Grant No. 10044723).

**Author contributions**

X.Q. and H.Y. conceived the experiments. X.Q. and K.N. carried out experiments. X.Q., P.D., and K-S.L. did modeling. X.Q., K.L., H.L., and H.Y. wrote the manuscript. All authors discussed the data and the results, and commented the manuscript. H.Y. supervised the project.

**Additional information**

**Supplementary Information** accompanies this paper on www.nature.com/scientificreports

**Competing financial interests:** The authors declare no competing financial interests.

Figure captions

**Fig. 1. Device schematic, magnetic property and current-induced switching of Ta/CoFeB/MgO nanowire. a,** Scanning electron micrograph of the device. **b**, Measurement geometry in the Cartesian coordinate system. **c**, Hall resistance $R_H$ as a function of $H$ applied to out-of-plane direction at 300 K. **d**, $R_H$ as a function of current with a 400 Oe magnetic field applied along the current direction at 300 K.

**Fig. 2. Schematics of two measurement schemes and current induced spin-orbit effective fields. a, b,** Directions of $H_L$ and $H_T$ with $H$ along and transverse to $I_{ac}$. First and second harmonic loops under these two schemes are shown in **c, d**, measured at $\theta_H = 4°$ and 300 K. **e** and **f** show the fitting of second harmonic voltage curves. The fitted $H_L$ and $H_T$ values are indicated in the graphs.

**Fig. 3. Angular dependence of current induced spin-orbit effective fields.** $H_L$ (**a**) and $H_T$ (**b**) vs. the magnetization angle $\theta$, measured at $\theta_H = 4°$ and 300 K.

**Fig. 4. Temperature dependence of current induced spin-orbit effective fields. a,** $H_L$ and $H_T$ (assuming constant values independent of magnetization direction) at $\theta_H = 4°$ at different temperatures. **b,** Comparison of $|1/I_S|$ (with 400 Oe assisted field along current direction) and $|H_L/H_{an}|$ at different temperatures. The inset shows the temperature dependence of $|H_T/H_{an}|$. **c**, $H_L$ vs. $\theta$ at different temperatures. **d**, $H_T$ vs. $\theta$ at different temperatures. The insets in **c, d** show the fits of second harmonic voltage curves by assuming angle-independent effective fields at 100 K, with the parameters of $H_L = -4.85$ Oe, $H_T = -7.50$ Oe, and $H_{an} = 9800$ Oe.

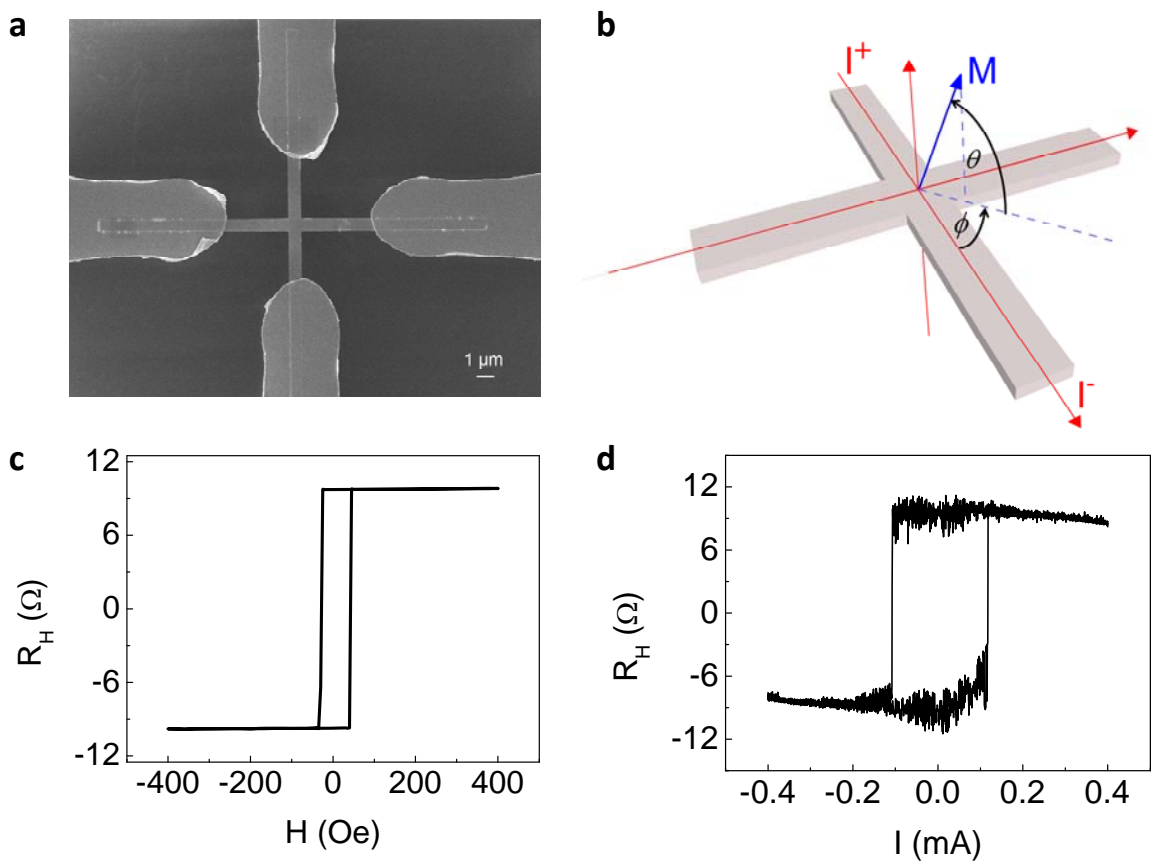

Fig. 1

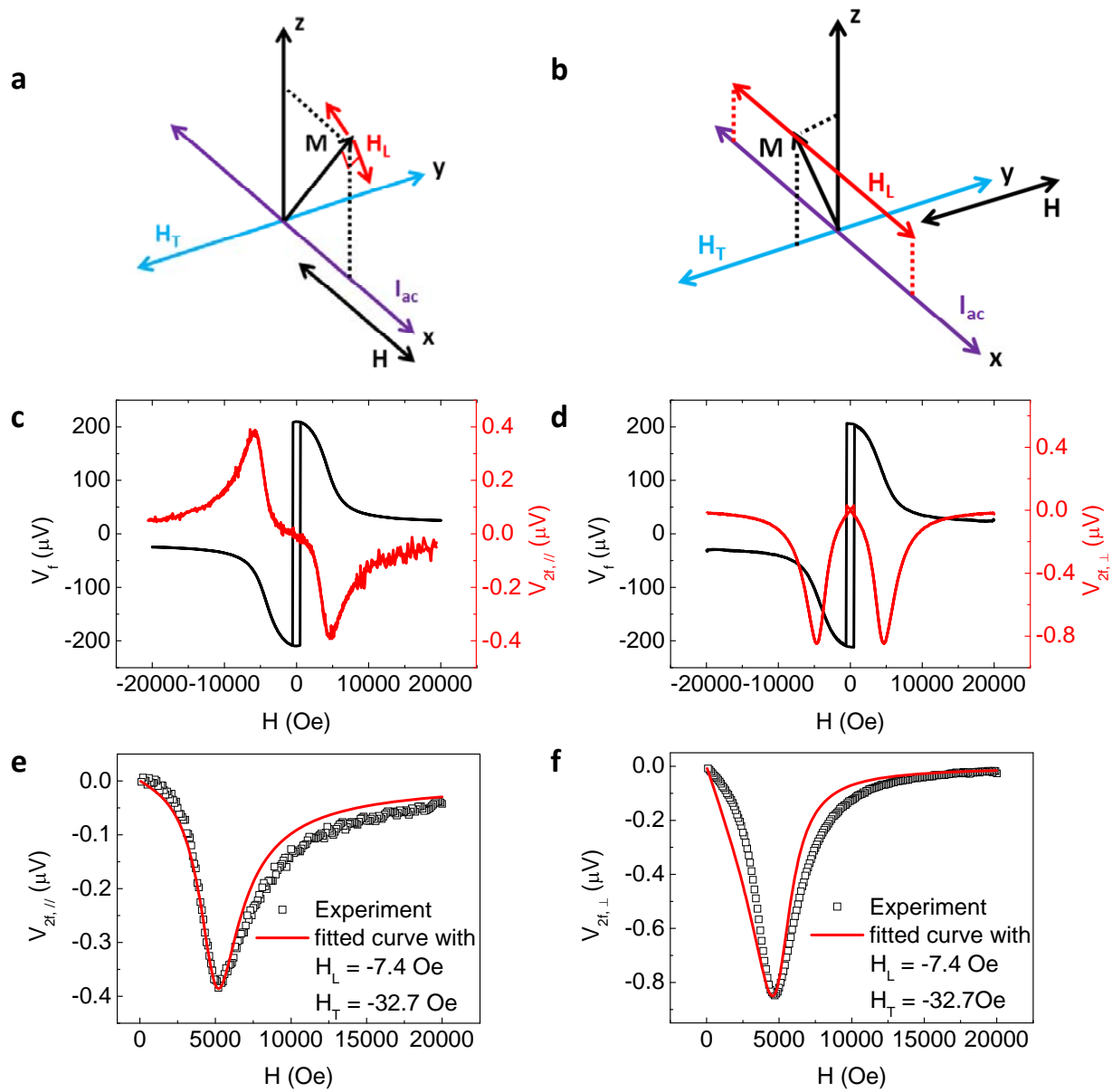

Fig. 2

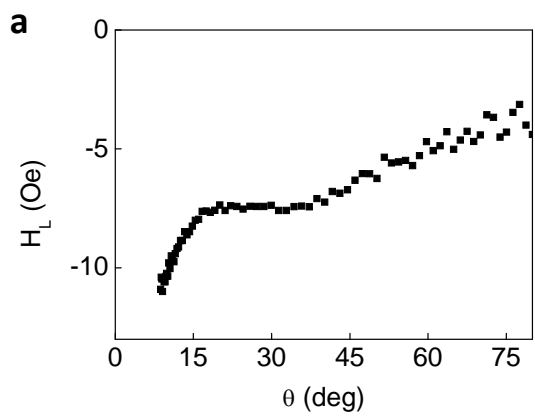 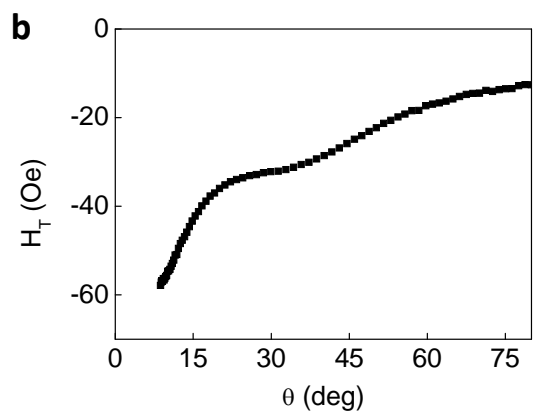

Fig. 3

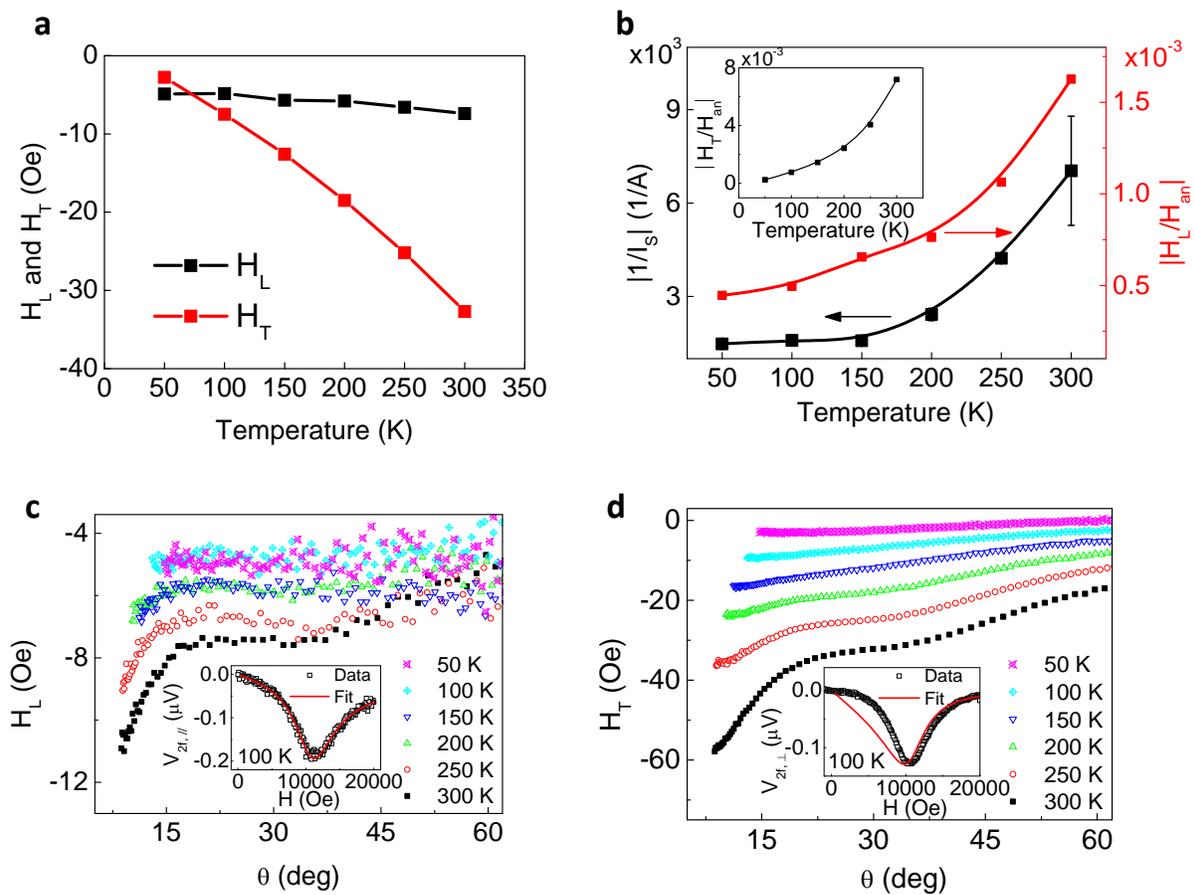

Fig. 4